# COMMENTS ON THE "GENERALIZED" KJLN KEY EXCHANGER WITH ARBITRARY RESISTORS: POWER, IMPEDANCE, SECURITY

SHAHRIAR FERDOUS, CHRISTIANA CHAMON, AND LASZLO B. KISH [1]

*Department of Electrical and Computer Engineering, Texas A&M University, TAMUS 3128, College Station, TX 77841-3128, USA*

*ferdous.shahriar@tamu.edu , cschamon@tamu.edu , laszlokish@tamu.edu*

In (Nature) Science Report 5 (2015) 13653, Vadai, Mingesz and Gingl (VMG) introduce a new Kirchhoff-law-Johnson-noise (KLJN) secure key exchanger that operates with 4 arbitrary resistors (instead of 2 arbitrary resistance values forming 2 identical resistor pairs in the original system). They state that in this new, VMG-KLJN, non-equilibrium system with nonzero power flow, the security during the exchange of the two (HL and LH) bit values is as strong as in the original KLJN scheme. Moreover, they claim that, at practical conditions, their VMG-KLJN protocol "supports more robust protection against attacks". First, we investigate the power flow and thermal equilibrium issues of the VMG-KLJN system with 4 arbitrary resistors. Then we introduce a new KLJN protocol that allows the arbitrary choice of 3 resistors from the 4, while it still operates with zero power flow during the exchange of single bits by utilizing a specific value of the 4th resistor and a binary temperature set for the exchanged (HL and LH) bit values. Then we show that, in general, the KLJN schemes with more than 2 arbitrary resistors (including our new protocol mentioned above) are prone to 4 new passive attacks utilizing the parasitic capacitance and inductance in the cable, while the original KLJN scheme is naturally immune against these new attacks. The core of the security vulnerability exploited by these attacks is the different line resistances in the HL and LH cases. Therefore, on the contrary of the statement and claim cited above, the practical VMG-KLJN system is less secure than the original KLJN scheme. We introduce another 2, modified, non-equilibrium KLJN systems to eliminate the vulnerability against some - *but not all* - of these attacks. However the price for that is the loss of arbitrariness of the selection of the 4th resistor and the information leak still remains greater than zero.

## 1. Introduction

Information theoretic (unconditional) security [1-4] is robust against arbitrarily high computation power, measurement accuracy and speed of the eavesdropper (Eve). Unconditional, hardware-based security is offered by quantum key distribution (QKD) [4-40] and its statistical-physical competitor, the Kirchhoff-law-Johnson-noise (KLJN) secure key exchanger [41-94]. The KLJN system operates with 4 arbitrary resistors, see Figure 1, which form two identical resistor pairs (each pair is of $R_L$ and $R_H$) at communicators A (Alice) and B (Bob), respectively. The resistance $R_L$ is less than the resistance $R_H$. At the beginning of each bit exchange period, Alice and Bob randomly and independently choose one of these resistors and connect them to the wire line for the whole period.

---

[1] Corresponding Author

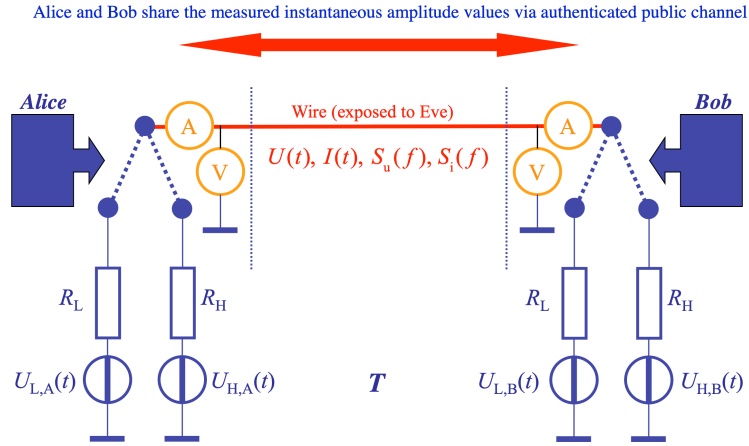

**Fig. 1.** The KLJN secure key exchanger scheme [3]. At the beginning of the bit exchange period, one of the resistors is randomly selected and connected at each end. The voltage generators represent the thermal noise of the resistors or much higher noise temperatures emulated by external, independent, Gaussian noise generators. The homogeneous temperature $T$ in the system guarantees that the LH (Alice $R_L$, Bob $R_H$) and HL (Alice $R_H$, Bob $R_L$) resistor connections provide the same mean-square voltage and current, and the related spectra, in the wire. For defense against various active (invasive) attacks, Alice and Bob are measuring the instantaneous voltage and current amplitudes and exchange these data via an authenticated public channel.

The key exchanger operates with the thermal noise of the resistors that can be emulated by external Gaussian voltage noise generators. The homogeneous temperature $T$ guarantees that, in the quasi-static frequency limit where the system must operate [3], the LH (Alice $R_L$, Bob $R_H$) and HL (Alice $R_H$, Bob $R_L$) resistor connections provide the same mean-square voltage and mean-square current in the wire. These are secure levels because the eavesdropper (Eve) knows that the situation is either HL or LH but she is uncertain which one. This is a 1-bit uncertainty (information entropy) for Eve.

On the other hand, when Alice detects that the noise is at the secure level, she knows that Bob has connected a different resistor than what she did. Thus Alice knows Bob's connected resistance value. The same argumentation works for Bob. A random bit generation and secure bit exchange takes place.

Alice and Bob publicly agree that which bit value (0 or 1) the HL situation represents. Naturally, LH represents the opposite bit value.

The unconditional security of the KLJN scheme is provided by the Second Law of Thermodynamics, which requires thermal equilibrium (homogeneous temperature) for the system. Nonzero energy flow in the wire implies nonzero information leak [81,82], that is, less than 1 bit information entropy for Eve (see also [47,48,84]).

This claim about the thermal equilibrium requirement, as the Second Law foundation of the security, was seriously challenged by Vadai-Mingesz-Gingl (VMG) [48], who claimed that their impressive VMG-KLJN system can operate with 4 different resistance values and still provide the same perfect security level, even though inhomogeneous temperature and nonzero power flow are required for that security.

Subsequently, Kish, inspired by the VMG-KLJN system, introduced the random-resistor-random-temperature (RRRT) KLJN protocol [49] that is also claimed to be as secure as the original KLJN key exchanger while the power flow is typically non-zero.

We have been investigating these claims and, in various ways, we concluded that these claims are incorrect because there are new attack types that work against these new schemes but not against the original KLJN system, which is a proof that the new non-equilibrium schemes are not as secure as the original KLJN system. In other words, thermal equilibrium remains the foundation of the security.

This is the first paper about our results, where we show that, under practical conditions with nonzero cable capacitance and inductance, the VMG-KLJN scheme is vulnerable to certain attacks while the original KLJN scheme is resistant against the same attacks. In other words, the VMG-KLJN (and probably the RRRT-KLJN) system is less secure than the original KLJN protocol. We also show that some of these vulnerabilities can be fixed by yet another new protocol that we introduce here. However at least one of these vulnerabilities will always remain, thus the information leak is never mathematically zero.

### 1.1. *The VMG-KLJN scheme*

In the KLJN protocol, the net power flow $P_{AB}$ between Alice and Bob is zero because their resistors have the same temperature. The noise spectrum of the voltage $U$ and current $I$ in the wire, $S_u$ and $S_i$, respectively, are given by the Johnson formulas of thermal noise:

$$S_u(f) = 4kTR_p , \tag{1}$$

$$S_I(f) = \frac{4kT}{R_s} , \tag{2}$$

where k is the Boltzmann constant, and $R_p$ and $R_s$ are the parallel and serial resultant of the connected resistors, respectively. In the HL and LH cases the resultants are:

$$R_{pLH} = R_{pHL} = \frac{R_L R_H}{R_L + R_H} \tag{3}$$

$$R_{sLH} = R_{sHL} = R_L + R_H . \tag{4}$$

Equations 1-4 guarantee that the noise spectra and effective values, $U$ and $I$, of the voltage and current in the wire, respectively, are identical in the LH and HL cases in accordance with the perfect security of the scheme. In conclusion, the quantities that Eve can access with passive measurements satisfy the following equations that, together with Equations 3-4, form the pillars of security against passive attacks against the KLJN scheme:

$$U_{LH} = U_{HL} \tag{5}$$

$$I_{LH} = I_{HL} \tag{6}$$

$$P_{LH} = P_{HL} = 0 \tag{7}$$

Vadai, et al, [48] made an impressive generalization attempt: they assumed that the four resistors are different and asked the question if the security can be maintained by a proper choice of (different) temperatures of these resistors.

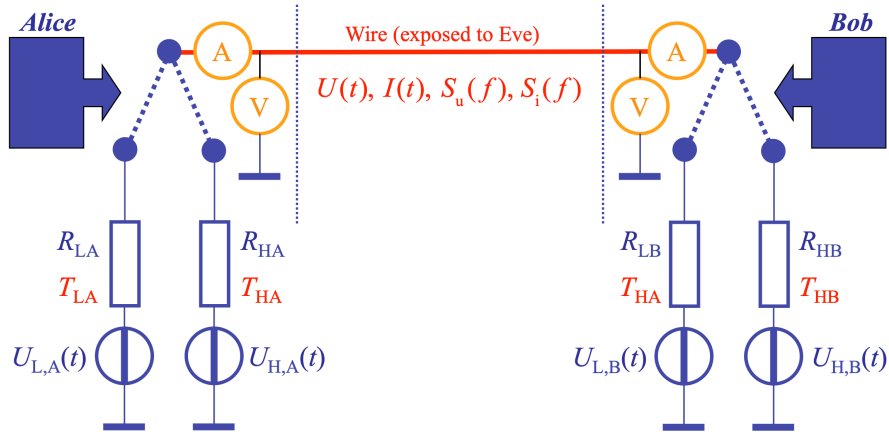

**Fig. 2.** The core of the VMG-KLJN secure key exchanger scheme. The four resistors are different and can be freely chosen (though not arbitrarily because of certain unphysical solutions). One temperature is freely chosen and the other 3 temperatures depend on the resistor values and can be deducted by the VMG-equations 9-11, see Equations 13-15 below.

In search for their solution, they used Equations 5 and 6 and they modified Equation 7 by removing the zero power flow conditions:

$$P_{LH} = P_{HL} \tag{8}$$

VMG obtained the following solutions for the voltages of the noise generators of the resistors, where the voltage $U_{LA}$ of the resistor $R_{LA}$ is freely chosen:

$$U_{HB}^2 = U_{LA}^2 \frac{R_{LB}(R_{HA} + R_{HB}) - R_{HA}R_{HB} - R_{HB}^2}{R_{LA}^2 + R_{LB}(R_{LA} - R_{HA}) - R_{HA}R_{LA}} = 4kT_{HB}R_{HB}B \quad , \tag{9}$$

$$U_{HA}^2 = U_{LA}^2 \frac{R_{LB}(R_{HA} + R_{HB}) + R_{HA}R_{HB} + R_{HA}^2}{R_{LA}^2 + R_{LB}(R_{LA} + R_{HB}) + R_{HB}R_{LA}} = 4kT_{HA}R_{HA}B \ , \tag{10}$$

$$U_{LB}^2 = U_{LA}^2 \frac{R_{LB}(R_{HA} - R_{HB}) - R_{HA}R_{HB} + R_{LB}^2}{R_{LA}^2 + R_{LA}(R_{HB} - R_{HA}) - R_{HA}R_{HB}} = 4kT_{LB}R_{LB}B \ . \tag{11}$$

Here we expanded the VMG equations by introducing (on the right hand side) the temperatures of the resistors, where *B* is the noise bandwidth of the generators, which is identical for all resistors, and the required temperatures of the resistors shown above are determined by the Johnson-Nyquist formula:

$$U^2 = 4kTRB \ . \tag{12}$$

Thus, from Equations 9-11, the temperatures are:

$$T_{HB} = \frac{R_{LA}}{R_{HB}} T_{LA} \frac{R_{LB}(R_{HA} + R_{HB}) - R_{HA}R_{HB} - R_{HB}^2}{R_{LA}^2 + R_{LB}(R_{LA} - R_{HA}) - R_{HA}R_{LA}} \ , \tag{13}$$

$$T_{HA} = \frac{R_{LA}}{R_{HA}} T_{LA} \frac{R_{LB}(R_{HA} + R_{HB}) + R_{HA}R_{HB} + R_{HA}^2}{R_{LA}^2 + R_{LB}(R_{LA} + R_{HB}) + R_{HB}R_{LA}} \ , \tag{14}$$

$$T_{LB} = \frac{R_{LA}}{R_{LB}} T_{LA} \frac{R_{LB}(R_{HA} - R_{HB}) - R_{HA}R_{HB} + R_{LB}^2}{R_{LA}^2 + R_{LA}(R_{HB} - R_{HA}) - R_{HA}R_{HB}} \ , \tag{15}$$

where $T_{LA}$ is the temperature of resistor $R_{LA}$.

Note, we have to correct a statement by VMG [48]: they write that the VMG scheme is essential because the resistance inaccuracies due to tolerance or temperature dependence or on any other parasitic resistance in the loop cannot be eliminated. Temperature dependence can be made immeasurably small by manganine or constantan resistors and/or temperature stabilization. Inaccuracies due to tolerance can be nullified by high-precision resistors. Aging of resistors, which is usually negligible, can also be eliminated by proper choice of materials, or resistance control that would require additional hardware overhead. The real advantage of the VMG scheme would appear with inexpensive versions of chip technology where resistance accuracy and temperature stability are poor.

We believe, the VMG-KLJN scheme [48] has strong potential applications for chip technology even if its security level is reduced compared to the original KLJN protocol. It is a security level that is sufficient for many practical applications but is reduced due to the deviation from the original KLJN. The rest of this paper, and another work in progress show further details about the superiority of the security of the original KLJN protocol.

*S. Ferdous, et al. Impedance attacks against the VMG-KLJN key exchange scheme*

## 2. A new, zero-power KLJN scheme with 4 different resistances

The power flow from Alice and Bob can be positive or negative. That naturally raises the question if the power flow can be zero when the four resistors are different while the *VMG-level of security* is still maintained. Note, here we mean *VMG-level of security* as a limited security compared to that of the original KLJN scheme.

The answer to the above question is *yes* and it comes from the solution of the VMG problem expanded by the zero-power constraint. The net power flow from Alice to Bob in the HL and LH cases is given by the superposition theorem as follows:

$$P_{HL} = \frac{R_{LB}U_{HA}^2 - R_{HA}U_{LB}^2}{\left(R_{LB} + R_{HA}\right)^2} \tag{16}$$

$$P_{LH} = \frac{R_{HB}U_{LA}^2 - R_{LA}U_{HB}^2}{\left(R_{LA} + R_{HB}\right)^2} \tag{17}$$

The solution of the system of Equations 7, 9-11, 16 and 17 for the resistances yields that the condition of zero power is that the geometrical means of the connected resistors in the LH and HL situations are equal. In other words, when we choose three resistors freely, the fourth one is determined by the condition of zero power flow. For example, with chosen $R_{HB}$, $R_{LA}$ and $R_{HA}$, we get:

$$R_{LB} = \frac{R_{HB}R_{LA}}{R_{HA}} \tag{18}$$

It is important to note that Equation 18 *does not guarantee thermal equilibrium* in the resistor system; it guarantees zero power flow only! It provides thermal equilibrium, that is, identical LH and HL temperatures, *solely* in the situation of the original KLJN scheme, where

$$R_{LB} = R_{LA} = R_L \quad \text{and} \quad R_{HB} = R_{HA} = R_H \quad . \tag{19}$$

That means, in the general VMG-KLJN situation, where the four resistors are different, only the temperature of the *actually connected resistors* will be identical but this temperature (see Equation 12-15) will be different in the LH and HL cases. In other words, Alice will have *inhomogeneous temperature* in her system, similarly to Bob. These systems are in *steady-state, out of equilibrium*, not in thermal equilibrium state. The implication is that their security will be reduced compared to the original security of the KLJN protocol, as it will be shown in the rest of the paper.

In Table 1, three different examples for zero power flow during secure bit exchange are shown. Case A is an original KJLN scheme with thermal equilibrium while cases B and C are situations with zero power flow but out of equilibrium.

|  | **A** | **B** | **C** |
|---|---|---|---|
| $R_{HA}$ [Ohm] | 9000 | 9000 | 9000 |
| $R_{LB}$ [Ohm] | 1000 | 1000 | 1000 |
| $R_{LA}$ [Ohm] | 1000 | 500 | 2000 |
| $R_{HB}$ [Ohm] | 9000 | 18000 | 4500 |
| $U_{HA}$ [V] | 3 | 3.12 | 2.63 |
| $U_{LB}$ [V] | 1 | 1.04 | 0.877 |
| $U_{LA}$ [V] | 1 | 1 | 1 |
| $U_{HB}$ [V] | 3 | 6 | 1.5 |
| $T_{HA}$ [K] | $1.81 \times 10^{16}$ | $1.96 \times 10^{16}$ | $1.39 \times 10^{16}$ |
| $T_{LB}$ [K] | $1.81 \times 10^{16}$ | $1.96 \times 10^{16}$ | $1.39 \times 10^{16}$ |
| $T_{LA}$ [K] | $1.81 \times 10^{16}$ | $3.62 \times 10^{16}$ | $9.06 \times 10^{15}$ |
| $T_{HB}$ [K] | $1.81 \times 10^{16}$ | $3.62 \times 10^{16}$ | $9.06 \times 10^{15}$ |
| $U_{HL}$ [V] | 0.948 | 0.986 | 0.832 |
| $U_{LH}$ [V] | 0.948 | 0.986 | 0.832 |
| $I_{HL}$ [A] | $3.16 \times 10^{-4}$ | $3.29 \times 10^{-4}$ | $2.77 \times 10^{-4}$ |
| $I_{LH}$ [A] | $3.16 \times 10^{-4}$ | $3.29 \times 10^{-4}$ | $2.77 \times 10^{-4}$ |
| $P_{HL}$ [W] | 0 | 0 | 0 |
| $P_{LH}$ [W] | 0 | 0 | 0 |

Table 1. Zero power flow KLJN schemes where the connected resistors have identical temperatures during the secure bit exchange thus the power flow is zero. For the temperature data (see Equation 12), noise bandwidth $B$=1000 Hz is assumed. First column: classical KLJN protocol with identical resistor pairs and identical temperatures of both (HL and LH) resistor pairs: the same choice as in the KLJN example in [48]. Second and third columns: the HL and LH resistor pairs are different which implies that their temperatures must be different. The power flow is still zero because the temperature within each connected resistor pair is homogeneous even though the two pairs (HL and LH) have different temperatures. Thus, the whole KLJN system has binary temperature indicating that the system is a steady-state one, out of equilibrium.

### 3. Information leak in the VMG protocol

Below we describe simple attacks to utilize practical security weaknesses of the VMG-KLJN scheme and later we offer a partial solution of these problems that can be sufficient for some practical applications.

*3.1 Security vulnerability due to bit-dependent impedances*

*3.1.1. Impedances in the VMG-KLJN scheme*

In the original KLJN scheme, the parallel resultant $R_p$ and serial resultant $R_s$ of the connected resistances are identical in the LH and HL cases. However, in the VMG-KLJN system these conditions are not guaranteed. Moreover, even if the $R_p$ ($R_s$) values are designed to be identical in the LH and HL cases, see Section 4.2, the $R_s$ ($R_p$) values will be different unless the scheme is the original KLJN system. In other words, if the parallel resultants match, the serial resultants do not and *vice versa*. If Eve can do a *passive* measurement of the $R_p$ and $R_s$ values, she can crack the VMG-KLJN scheme. *Active* measurements (such as current injection) would easily permit this attack but passive measurements are less easy. Below, in Sections 3.1.3 and 3.2, we show two passive attacks (utilizing crossover frequency and noise temperature) for the cases when cable capacitance and inductances are not negligible.

Instead of Equations 3 and 4, the parallel and serial resultant resistances in the LH and HL cases of the VMG-KLJN scheme are as follows:

$$R_{pLH} = \frac{R_{LA} R_{HB}}{R_{LA} + R_{HB}} \,, \tag{20}$$

$$R_{pHL} = \frac{R_{HA} R_{LB}}{R_{LB} + R_{HA}} \,, \tag{21}$$

$$R_{sLH} = R_{LA} + R_{HB} \,, \tag{22}$$

$$R_{sHL} = R_{HA} + R_{LB} \,. \tag{23}$$

When the cable capacitance $C_c$ is not negligible, it will form a first-order low-pass filter with the parallel resistance for the resultant voltage noise on the wire. Similarly, when the cable inductance $L_c$ is not negligible, it will form a first-order low-pass filter with the serial resistance for the resultant current noise in the wire.

*3.1.2. Noise spectra in the VMG-KLJN scheme*

The generic shape of the power density spectra S($f$) of voltage (current) on (in) the wire is Lorentzian [95] (see its linear plot in Figure 3):

$$S(f) = \frac{S(0)}{1 + f^2 / f_{cr}^2} \,, \tag{24}$$

where $S(0)$ is the spectrum in the zero frequency limit and $f_{cr}$ is the crossover frequency (pole frequency).

For the particular VMG-KLJN system operation, the values of above defined quantities will depend on the situation while the qualitative spectral properties remain the same.

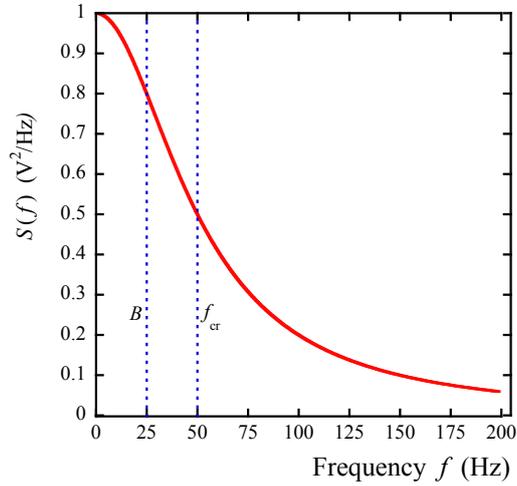

**Fig. 3.** Example: Linear plot of Lorentzian noise spectrum with crossover frequency $f_{cr}$=50Hz at infinite *noise bandwidth* [95] of the generator. As an illustration, a practical noise bandwidth $B$=25 Hz is also indicated ($B<<f_{cr}$), where the resultant spectrum would cut off to zero due to the cut-off in the noise generation.

Figures 4 and 5 show the resultant circuit situations of the noise voltage and noise current, respectively.

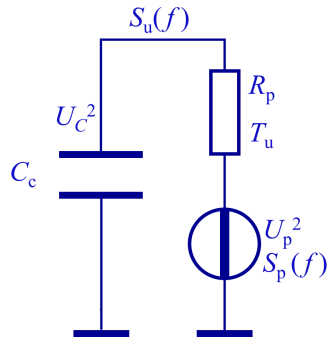

**Fig. 4.** Generic circuit model of the resultant parallel voltage generator of the circuitries in Figures 1, 2 and its internal resistance (parallel resultant $R_p$ of Alice's and Bob's connected resistances) with the cable capacitance shunt $C_c$. $U_p^2$ and $S_p$ are the resultant mean-square voltage and voltage noise spectrum of the *parallel circuit* of connected resistances at Alice and Bob's sides, respectively; and $S_u$ is the voltage noise spectrum modified by the cable capacitance. $U_C^2$ is the mean-square voltage noise on the cable. For the LH and HL cases, the corresponding resistance and spectrum values must be used.

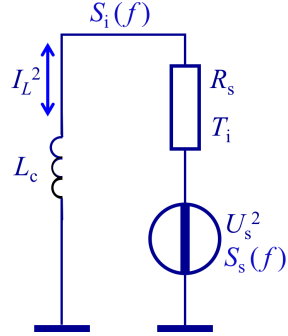

**Fig. 5.** Generic circuit model of the resultant serial voltage generator of the circuitries in Figures 1, 2 and its internal resistance (serial resultant $R_s$ of Alice's and Bob's connected resistances) with the serial cable inductance $L_c$. $U_s^2$ and $S_s$ are the resultant mean-square voltage and voltage noise spectrum of the *serial circuit* of the connected resistances at Alice's and Bob's sides, respectively; where $S_s$ is the sum of the voltage noise spectra of the serial resistors. $S_i$ is the spectrum of current noise in the wire. $I_L^2$ is the mean-square current noise in the cable. For the LH and HL cases, the corresponding resistor and spectrum values must be used.

The parameters for the respective Lorentzian spectra are discussed below. Due to the VMG-KLJN conditions (Equations 5 and 6), the zero-frequency voltage and current spectra are publicly known and they are identical in the LH and HL situations,

$$S_{uLH}(0) = S_{uHL}(0) = S_p(0) \quad \text{and} \quad S_{iLH}(0) = S_{iHL}(0) = \frac{S_s(0)}{R_s^2} , \tag{25}$$

because, in this limit, the impacts of $C_c$ and $L_c$ are nonexistent. The crossover frequencies of the Lorentzian spectra of voltage and current noises in the cable are given as:

$$f_{ucr} = \frac{1}{2\pi R_p C_c} \tag{26}$$

$$f_{icr} = \frac{1}{2\pi L_c / R_s} \tag{27}$$

Due to Equations 20-23, the crossover frequencies for the LH and HL situations will differ in an infinite number of situations, and even if they are identical for the voltage, they must be different for the current, and *vice versa*, see also Section 4.2.3

### 3.1.3. Crossover frequency attack against the VMG-KLJN scheme

Our first attack is based on the split of crossover frequencies in the LH and HL situations.

From Equation 24, the crossover frequency can be expressed as follows:

$$f_{cr} = \frac{f}{\sqrt{\frac{S(0)}{S(f)} - 1}} \quad , \tag{28}$$

where $S(0)$ is known (see Equation 25) and $S(f)$ is the *measured* spectrum at a chosen frequency $f$ within the noise bandwidth $B$. For a better statistics, $f_{cr}$ can be evaluated at multiple frequencies, and the final result can be obtained as an average (possibly properly weighted). The $f_{cr}$ measured in this way can be compared by the theoretical predictions for the LH and HL bit situations, and the secure LH or LH situation can be identified by Eve.

The theoretical values are as follows: for the voltage:

$$f_{ucrLH} = \frac{R_{LA} + R_{HB}}{2\pi R_{LA} R_{HB} C_c} \quad \text{and} \quad f_{ucrHL} = \frac{R_{HA} + R_{LB}}{2\pi R_{HA} R_{LB} C_c} \quad , \tag{29}$$

and for the current:

$$f_{icrLH} = \frac{R_{LA} + R_{HB}}{2\pi L_c} \quad \text{and} \quad f_{icrHL} = \frac{R_{HA} + R_{LB}}{2\pi L_c} \quad . \tag{30}$$

Equations 29 and 30 prove that the VMG-KLJN protocol does *not* guarantee that the LH and HL crossover frequencies are equal. Therefore the VMG-KLJN protocol allow a nonzero information leak in infinitely many situations, whenever:

$$f_{crHL} \neq f_{crLH} \tag{31}$$

for the voltage or the current or both. At least one of them will have a mismatch guaranteeing imperfect security against a simple one-point measurement on the wire.

It is important to note that these parasitic capacitance and inductance effects can also be utilized to extract information from the original KLJN scheme but these would require doing two-point measurements, probing both ends of the cable and comparing the data, such as in the cable resistance attacks [83, 84] or in the cable capacitance attack [86]. The major difference is that in the VMG-KLJN case, a single one-point measurement is enough, which indicates that the original KLJN scheme has higher level of security than the VMG scheme as there exists a new, simpler attack method that works against the VMG-KLJN scheme but it is ineffective against the original KLJN protocol.

Finally, Tables 2 and 3 show some practical data for the crossover frequencies. With the selected values, it is obvious that the crossover frequency of the current will provide much less (though nonzero) information leak indicating that similar choices could be a practically useful way to reduce the information leak.

| **HL** | **LH** | **Bit resistance (parallel) [kOhm]** | **Crossover frequencies of voltage noise spectra** |
|---|---|---|---|

| $R_{HA}$ [kOhm] | $R_{LB}$ [kOhm] | $R_{LA}$ [kOhm] | $R_{HB}$ [kOhm] | $R_{pHL}$ | $R_{pLH}$ | $f_{ucrHL}$ [Hz] | $f_{ucrLH}$ [Hz] |
|---|---|---|---|---|---|---|---|
| 9 | 1 | 1 | 9 | 0.9 | 0.9 | 884 | 884 |
| 10 | 5 | 1 | 9 | 3.33 | 0.9 | 239 | 884 |
| 5 | 5 | 1 | 9 | 2.5 | 0.9 | 318 | 884 |

Table 2. Crossover frequency based information leak due to the cable capacitance. The data are based on VMG's Table 2 [48] with $U_{LA} = 1$ V and KLJN bandwidth $B = 1$ kHz. A cable with 2000 meter length and 100 picoFarad/meter specific capacitance [91] is assumed. The first case is a classical KLJN situation thus the HL and LH crossover frequencies are naturally identical. However the HL and LH crossover frequencies are different in the second and the third cases indicating a nonzero information leak about the exchanged bit, in each case.

| HL | | LH | | Serial bit resistance [kOhm] | | Crossover frequencies of current noise spectra | |
|---|---|---|---|---|---|---|---|
| $R_{HA}$ [kOhm] | $R_{LB}$ [kOhm] | $R_{LA}$ [kOhm] | $R_{HB}$ [kOhm] | $R_{sHL}$ | $R_{sLH}$ | $f_{icrHL}$ [Hz] | $f_{icrLH}$ [Hz] |
| 9 | 1 | 1 | 9 | 10 | 10 | $1.14 \times 10^6$ | $1.14 \times 10^6$ |
| 10 | 5 | 1 | 9 | 15 | 10 | $1.71 \times 10^6$ | $1.14 \times 10^6$ |
| 5 | 5 | 1 | 9 | 10 | 10 | $1.14 \times 10^6$ | $1.14 \times 10^6$ |

Table 3. Crossover frequency based information leak due to the cable inductance. The data are based on VMG's Table 2 with $U_{LA} = 1$ V and KLJN bandwidth $B = 1$ kHz. A cable with 2000 meter length and 0.7 microH/m specific inductance [91] is assumed. The first case is a classical KLJN situation where the crossover frequencies are naturally identical. The crossover frequencies are different in the second case but identical in the third case.

*3.2 Security vulnerability due to bit-dependent noise temperature*

The *original* KLJN system guarantees that the effective noise temperatures of the wire line as a generator – for both the voltage ($T_u$) and current ($T_i$) (see Figures 4-5) – are identical in the LH and HL bit situations. In other words, the bit-temperatures $T_{LH}$ and $T_{HL}$ satisfy:

$$T_{LH} = T_{HL} = T, \tag{32}$$

and these temperatures are the same for both the current and the voltage. However, in the VMG-KLJN system, these conditions are not guaranteed. Moreover, even if one class of the conditions is satisfied by proper design (see our new KLJN versions in Section 4), the other one will be violated. For example, if the voltage noise temperatures match, the current noise temperatures cannot and *vice versa*.

Thus, another physical way to extract information about the shared bit is the exploitation of the varying effective noise temperature of the line (see Figures 4-5).

*3.2.1 The thermodynamical picture behind the bit-dependent noise temperature*

The physical reason for the information leak is Boltzmann's energy equipartition theorem – with its $kT/2$ mean energy per thermal degree of freedom – implying different mean thermal energies in the cable capacitance and inductance for different temperatures. As it is well known, in thermal equilibrium, the mean-square noise voltage on a shunt capacitance *C* (see Figure 4) and the mean-square noise current through a shunt inductance *L* (see Figure 5) fed by the thermal noise of a resistive network are given as:

$$U_C^2 = \int_0^\infty \frac{S_\mathrm{u}(0)}{1+f^2/f_\mathrm{ucr}^2} = \frac{kT}{C} \ , \qquad (33)$$

$$I_L^2 = \int_0^\infty \frac{S_\mathrm{i}(0)}{1+f^2/f_\mathrm{icr}^2} = \frac{kT}{L} \ , \qquad (34)$$

where $U_C$ and $I_L$ are the effective (rms) values, and an infinite noise generator bandwidth *B* is supposed.

*3.2.2 Noise temperature attack against the VMG-KLJN scheme*

The different effective temperatures in the HL and LH cases imply different mean-square quantities:

$$\text{If } T_{U\mathrm{HL}} \neq T_{U\mathrm{HL}} \text{ then } U_{C\mathrm{HL}}^2 \neq U_{C\mathrm{LH}}^2 \ , \qquad (35)$$

$$\text{If } T_{I\mathrm{HL}} \neq T_{I\mathrm{HL}} \text{ then } I_{L\mathrm{HL}}^2 \neq I_{L\mathrm{LH}}^2 \ . \qquad (36)$$

In the case of practical KLJN systems, the noise bandwidth *B* is not infinite but it is chosen to approach the quasi-static situation:

$$B \ll f_{\mathrm{cr,min}} \ , \qquad (37)$$

where $f_{\mathrm{cr,min}}$ is the smallest of all the four crossover frequencies defined by Equations 29 and 30. However it is obvious from Equations 24, 29, 30, 33 and 34, and from Figure 3 that Relations 35 and 36 remain valid for *any nonzero bandwidth B*. The Lorentzian spectra in zero frequency limit are identical in the LH and HL states due to VMG conditions (Equations 5, 6) and the varying crossover frequencies (Equations 29, 30) yield varying decays and mean-square values within the *B* bandwidth.

In conclusion, both Relations 35 and 36 serve as a temperature-based attack tool. Finally, we give a few additional remarks below:

a) The voltages and currents in the above equations are measurable quantities in the cable, and due to the nonzero parasitic capacitances and inductances, for nonzero *B*, they are always less than VMG conditions Equations 5, 6 dictate:

$$U_{CHL}^2 < U_{HL}^2 \ , \ U_{CLH}^2 < U_{LH}^2 \ , \ I_{LHL}^2 < I_{HL}^2 \ , \ \text{and} \ I_{LLH}^2 < I_{LH}^2 \ . \tag{38}$$

b) For the examples about temperatures shown in this paper, we used the following protocol (based on Equations 39 and 40). The effective temperatures are related to the cable voltages and the parallel resultants of driving resistances; and the cable currents and the serial resultants of driving resistances. In the limit of negligible capacitances and inductances, the noise is white within the noise bandwidth *B* thus:

$$T_{uHL} = \frac{U_{HL}^2}{4kBR_{pHL}} = \frac{U_{HL}^2(R_{HA} + R_{LB})}{4kBR_{HA}R_{LB}} \quad \text{and} \quad T_{uLH} = \frac{U_{LH}^2}{4kBR_{pLH}} = \frac{U_{HL}^2(R_{LA} + R_{HB})}{4kBR_{LA}R_{HB}} \ , \tag{39}$$

and

$$T_{iHL} = \frac{I_{HL}^2 R_{sHL}}{4kB} = \frac{I_{HL}^2(R_{HA} + R_{LB})}{4kB} \quad \text{and} \quad T_{iLH} = \frac{I_{LH}^2 R_{sLH}}{4kB} = \frac{I_{HL}^2(R_{LA} + R_{HB})}{4kB} \ . \tag{40}$$

As an illustration, Tables 4 (voltage) and 5 (current) show the bit temperatures in three different situations. The data are based on VMG's Table 2 with $U_{LA} = 1$ V and KLJN bandwidth $B = 1$ kHz. The first situation is the *original* KLJN scheme where both the HL and LH temperatures are identical. Also, the voltage-based and the current-based temperatures are also identical while in the VMG-KLJN situations this can never be the case.

| HL | | LH | | Bit resistance (parallel) [kOhm] | | Channel noise voltage temperature | |
|---|---|---|---|---|---|---|---|
| $R_{HA}$ [kOhm] | $R_{LB}$ [kOhm] | $R_{LA}$ [kOhm] | $R_{HB}$ [kOhm] | $R_{pHL}$ | $R_{pLH}$ | $T_{uHL}$ [K] | $T_{uLH}$ [K] |
| 9 | 1 | 1 | 9 | 0.9 | 0.9 | 1.81 x 10$^{16}$ | 1.81 x 10$^{16}$ |
| 10 | 5 | 1 | 9 | 3.33 | 0.9 | 4.48 x 10$^{15}$ | 1.66 x 10$^{16}$ |
| 5 | 5 | 1 | 9 | 2.5 | 0.9 | 6.04 x 10$^{15}$ | 1.65 x 10$^{16}$ |

Table 4. Bit temperature based information leak due to the mean-square voltage on the cable capacitance. The data are based on VMG's Table 2 with $U_{LA}$ = 1 V and KLJN bandwidth $B$ = 1 kHz. The first case is a classical KLJN situation thus the HL and LH voltage noise temperatures are naturally identical. However the HL and LH voltage noise temperatures are different in the second and the third cases indicating a nonzero information leak about the exchanged bit, in each case.

| HL | | LH | | Serial bit resistance [kOhm] | | Channel noise current temperature | |
|---|---|---|---|---|---|---|---|
| $R_{HA}$ [kOhm] | $R_{LB}$ [kOhm] | $R_{LA}$ [kOhm] | $R_{HB}$ [kOhm] | $R_{sHL}$ | $R_{sLH}$ | $T_{iHL}$ [K] | $T_{iLH}$ [K] |
| 9 | 1 | 1 | 9 | 10 | 10 | 1.81 x 10$^{16}$ | 1.81 x 10$^{16}$ |
| 10 | 5 | 1 | 9 | 15 | 10 | 6.54 x 10$^{15}$ | 4.36 x 10$^{15}$ |
| 5 | 5 | 1 | 9 | 10 | 10 | 6.04 x 10$^{15}$ | 6.04 x 10$^{15}$ |

Table 5. Bit temperature based information leak due to the mean-square current in the cable inductance. The data are based on VMG's Table 2 with $U_{LA}$ = 1 V and KLJN bandwidth $B$ = 1 kHz. The first case is a classical KLJN situation where the HL and LH current noise temperatures are naturally identical and they are also identical with the noise voltage temperature. However, the current noise temperatures are different in the second case and identical in the third case.

## 4. Defense methods to reduce the excess information leak of the VMG system

We show two defense methods to reduce the information leak due to crossover frequency and temperature variations between the LH and HL situations. Neither of these methods can reduce the information leak to zero during the attacks described in the former section. On the other hand, the original KLJN system has zero leak against these attacks.

i) Noise bandwidth reduction. However, the price is reduced key exchange speed.

ii) Impedance matching by properly choosing the fourth resistance. It does not reduce key exchange speed. This method guarantees identical crossover frequencies and temperatures but only for the voltages or for the currents.

### *4.1 Reducing the noise bandwidth B of generators*

When the noise bandwidth $B$ is not infinite and the capacitances and inductances are not negligible, Equations 33 and 34 change to:

$$U_C^2(B) = \int_0^B \frac{S_U(0)}{1+f^2/f_{Ucr}^2} = S_U(0) f_{Ucr} \tan^{-1}\left(\frac{B}{f_{Ucr}}\right) , \quad (41)$$

$$I_L^2(B) = \int_0^B \frac{S_I(0)}{1+f^2/f_{Icr}^2} = S_I(0) f_{Icr} \tan^{-1}\left(\frac{B}{f_{Icr}}\right) . \tag{42}$$

If $B \to 0$, $U_C^2(B) \to B S_U(0)$ and $I_L^2(B) \to B S_I(0)$ , (43)

thus the information leak converges to zero because, due to the VMG conditions,

$$S_{U\mathrm{HL}}(0) = S_{U\mathrm{LH}}(0) \text{ and } S_{I\mathrm{HL}}(0) = S_{I\mathrm{LH}}(0) \tag{44}$$

thus

$$U_{C\mathrm{HL}}^2 \simeq U_{C\mathrm{LH}}^2 \text{ and } I_{L\mathrm{HL}}^2 \cong I_{L\mathrm{LH}}^2 \tag{45}$$

*4.2 Selecting a resistor set with identical bit-impedances and identical bit-temperatures*

Impedance matching of the LH and HL states by properly choosing the fourth resistance eliminates one of the information leaks described above: either the one in the voltage or the one in the current, that is, the related leaks through the crossover frequencies and noise temperatures.

*4.2.1 Identical parallel LH and HL bit resistances: eliminating information leak in the voltage*

Due to Equations 5, 6, 29 and 39, when the parallel resultant resistances in the LH and HL cases are identical, then both the corresponding crossover frequencies and noise temperatures will be identical, too. To satisfy the equation

$$R_{p\mathrm{HL}} = R_{p\mathrm{LH}} , \tag{46}$$

the solution of Equations 20, 21 and 46 for $R_{\mathrm{HB}}$ yields the condition to eliminate the information leak due to the mismatch in voltage-based crossover frequencies and noise temperatures:

$$R_{\mathrm{HB}} = \frac{R_{\mathrm{HA}} R_{\mathrm{LA}} R_{\mathrm{LB}}}{R_{\mathrm{HA}} R_{\mathrm{LA}} - R_{\mathrm{HA}} R_{\mathrm{LB}} + R_{\mathrm{LA}} R_{\mathrm{LB}}} . \tag{47}$$

The obvious condition for passive circuitry and KLJN applicability (positive resistances) is:

$$R_{\mathrm{LA}} > R_{p\mathrm{LH}} \tag{48}$$

Table 6 shows a few examples where the VMG-KLJN scheme has identical LH and HL parallel resistances and immune against the crossover frequency and noise temperature attacks against the voltage on the wire.

|  | A | B | C |
|---|---|---|---|
| $R_{HA}$ [Ohm] | 2000 | 1000 | 10000 |
| $R_{LB}$ [Ohm] | 90 | 160 | 500 |
| $R_{LA}$ [Ohm] | 100 | 200 | 500 |
| $R_{HB}$ [Ohm] | 620.7 | 444.4 | 10000 |
| $R_{pHL}$ [Ohm] | 86.1 | 137.9 | 476.2 |
| $R_{pLH}$ [Ohm] | 86.1 | 137.9 | 476.2 |

Table 6. VMG-KLJN setup with identical LH and HL parallel resultant resistances determined by Equation 47. $R_{pHL}$ and $R_{pLH}$ are the parallel resultants of the connected resistances of Alice and Bob during the LH and HL secure bit situations. The case **C** is a classical KLJN arrangement.

*4.2.2 Identical serial LH and HL bit resistances: eliminating information leak in the current*

Due to Equations 5, 6, 30 and 40, when the serial resultant resistances in the LH and HL cases are identical, both the corresponding crossover frequencies and noise temperatures will be identical. To satisfy the equation

$$R_{sHL} = R_{sLH} , \qquad (49)$$

the solution of Equations 22, 23 and 49 for $R_{LB}$ yields the condition to eliminate the current-based crossover frequency and noise temperature leaks of information:

$$R_{LB} = R_{LA} + R_{HB} - R_{HA} . \qquad (50)$$

The obvious condition of passive circuitry and KLJN applicability (positive resistances) is:

$$R_{LA} + R_{HB} > R_{HA} . \qquad (51)$$

Table 7 shows a few examples where the VMG-KLJN scheme has identical LH and HL serial resistances and is immune against the crossover frequency and noise temperature attacks against the current in the wire.

|  | A | B | C |
|---|---|---|---|
| $R_{HA}$ [Ohm] | 2000 | 1000 | 10000 |
| $R_{LB}$ [Ohm] | 1000 | 500 | 5000 |
| $R_{LA}$ [Ohm] | 500 | 200 | 5000 |
| $R_{HB}$ [Ohm] | 2500 | 1300 | 10000 |
| $R_{sHL}$ [Ohm] | 3000 | 1500 | 15000 |
| $R_{sLH}$ [Ohm] | 3000 | 1500 | 15000 |

Table 7. VMG-KLJN setup with identical LH and HL *serial* resultant resistances. $R_{sHL}$ and $R_{sLH}$ are the serial resultants of the connected resistances of Alice and Bob during the LH and HL secure bit situations derived from Equation 50. The case **C** is a classical KLJN arrangement.

### 4.2.3 The impossibility to simultaneously preserve the parallel and serial resistances

In the VMG-KLJN scheme, *more than two resistances are different* among the four resistors. If the parallel bit resistance values are identical for HL and LH, they are different for the serial resistances. *Vice versa*, if the serial bit resistance values are identical for HL and LH, they are different for the parallel resistances.

The heuristic proof of this statement is obvious: If we have two different resistors $R_1$ and $R_2$, we can determine the parallel and serial resultants of these resistors, $R_p$ and $R_s$, respectively. If we change the resistance of $R_1$, we can properly change $R_2$ so that *either* the $R_p$ *or* the $R_s$ will remain the same as earlier. The only way to keep both of them preserved is to change $R_1$ to $R_2$ and to change $R_2$ to $R_1$, that is, by strictly following what the original KLJN scheme is doing when switching between LH and HL.

Finally, Table 8 shows examples with practical parameters to show the superiority of the classical KLJN system (column C) regarding security. The classical KLJN setting that guarantees that all the temperatures and crossovers are identical, thus they do not pose a security vulnerability.

| | **A** | **B** | **C** |
|---|---|---|---|
| $R_{HA}$ [Ohm] | 2000 | 100 | 10000 |
| $R_{LB}$ [Ohm] | 90 | 10 | 1000 |
| $R_{LA}$ [Ohm] | 100 | 50 | 1000 |
| $R_{HB}$ [Ohm] | 620.69 | 60 | 10000 |
| $U_{HA}$ [V] | 6.32 | 1.63 | 3.16 |
| $U_{LB}$ [V] | 0.9 | 1 | 1 |
| $U_{LA}$ [V] | 1 | 1 | 1 |
| $U_{HB}$ [V] | 1.96 | 1.63 | 3.16 |
| $R_{pHL}$ [Ohm] | 86.1 | 9.09 | 909 |
| $R_{pLH}$ [Ohm] | 86.1 | 27.3 | 909 |
| $R_{sHL}$ [Ohm] | 2090 | 110 | 11000 |
| $R_{sLH}$ [Ohm] | 720.69 | 110 | 11000 |
| $T_{uHL}$ [K] | $1.72 \times 10^{17}$ | $1.69 \times 10^{18}$ | $1.81 \times 10^{16}$ |
| $T_{uLH}$ [K] | $1.72 \times 10^{17}$ | $5.64 \times 10^{17}$ | $1.81 \times 10^{16}$ |
| $T_{iHL}$ [K] | $3.54 \times 10^{17}$ | $6.04 \times 10^{17}$ | $1.81 \times 10^{16}$ |
| $T_{iLH}$ [K] | $1.22 \times 10^{17}$ | $6.04 \times 10^{17}$ | $1.81 \times 10^{16}$ |
| $U_{HL}$ [V] | 0.903 | 0.921 | 0.953 |
| $U_{LH}$ [V] | 0.903 | 0.921 | 0.953 |
| $I_{HL}$ [A] | 0.00306 | 0.0174 | 0.000302 |
| $I_{LH}$ [A] | 0.00306 | 0.0174 | 0.000302 |
| $P_{HL}$ [W] | 0.000453 | -0.00606 | 0 |
| $P_{LH}$ [W] | 0.000453 | -0.00606 | 0 |
| $f_{ucrHL}$ | 9239 | 87535 | 875 |
| $f_{ucrLH}$ | 9239 | 29178 | 875 |
| $f_{icrHL}$ | 237596 | 12505 | $1.25 \times 10^6$ |
| $f_{icrLH}$ | 81930 | 12505 | $1.25 \times 10^6$ |

Table 8. Examples with practical parameters to show the superiority of the classical KLJN system regarding security. **A**: Compensation for identical HL/LH temperatures and crossover frequencies for the voltage, but not the current; **B**: Compensation for identical HL/LH temperatures and crossover frequencies for the current, but not the voltage; **C**: a classical KLJN setting that guarantees that all the temperatures and crossovers are identical, thus they do not pose a security vulnerability. Cable length 2000 meters, cable-specific capacitance 100 pF/m, cable-specific inductance 0.7 microH/m. Noise bandwidth (for temperature calculation) *B*=1 kHz.

**Conclusions**

The VMG-KLJN scheme working with arbitrary resistances (four arbitrary resistors) produces parallel and serial resistance values that depend on the bit situation LH and HL, respectively. This poses security leaks that can be exploited by Eve in various ways. We showed two passive attacks that can utilize these resultant resistance variations with single-point measurements on the wire in a physical case. It is important to note that classical KLJN scheme is totally immune against these attacks indicating that the original scheme with its *thermal equilibrium* situation has the ultimate security. The VMG-KLJN scheme is not a "generalized" scheme that can offer the same security level in the mathematical sense but a modified system that provides the free choice of all the resistors while allowing a non-zero information leak.

We showed three modified VMG-KLJN key exchange schemes. Two of these offered defense against either the voltage based or current based attacks but not against both. The third new key exchange system is working with zero power flow in a steady state, out of equilibrium. All these new schemes allow the free choice of three resistors and the fourth one is implied by our equations. All these schemes have non-zero information leak in the physical cases we have discussed.

The VMG-KLJN scheme and the three modified/improved systems described in this paper can still be used in practical applications because, with proper design, the information leak can be reduced to satisfactory levels. For example, by choosing high-enough resistances, the crossover frequency for the current noise can be made very high thus their differences will not count At the same time, by using our new method for equivalent parallel resistances (Equations 47, 48), the crossover frequencies for the voltage noise can be made identical.

Yet, at the fundamental level, the original KLJN scheme is the most secure and the other schemes discussed in this paper are compromised versions for practical advantages by introducing a *weaker vulnerability* (the ones we have shown here) to compensate for a *stronger accidental vulnerability* (deviations from the exact resistance values of the original KLJN system).